\documentclass[11pt,a4paper]{article}
\usepackage[utf8]{inputenc}
\usepackage{graphicx}
\usepackage{amsmath,amssymb,mathtools}
\usepackage{hyperref}
\hypersetup{colorlinks=true,linkcolor=blue,citecolor=blue,urlcolor=blue}
\usepackage{geometry}
\date{}

\title{{\bf Classical Polymerization of the Bianchi I Model with Deformed Poisson Structure}}
\author{Babak Vakili\thanks{email: ba.vakili@iau.ac.ir}\\\\{\small {\it Department of Physics, CT.C., Islamic Azad University, Tehran, Iran}}}

\begin{document}
\maketitle

\begin{abstract}
We study the dynamics of the Bianchi~I cosmological model in the presence of both polymer quantization 
effects and an exponential deformation of the Poisson algebra. Starting from the Hamiltonian formulation, we derive the polymer-deformed 
equations of motion and analyze their solutions for the contracting branch of the model. 
In contrast with the undeformed classical dynamics, the exponential deformation with suitable values of deformation parameters, 
produces a noticeably slower evolution of the volume variable and leads to a stabilization 
of the anisotropy parameters, which remain bounded throughout the evolution. 
No removal of the initial singularity is observed; however, the deformation significantly modifies the 
asymptotic behavior, offering a mechanism to suppress anisotropic shear near the singularity. 
Our results are illustrated through analytic solutions, highlighting the qualitative differences between the standard and the polymer--deformed Bianchi~I cosmology.
\vspace{5mm}\noindent\vspace{0.8mm}\newline Keywords: Bianchi I cosmology; Classical polymerization; Deformed Poisson brackets; Anisotropic models; Quantum cosmology
\end{abstract}

\section{Introduction}
One of the most persistent issues in classical cosmology is the presence of initial singularities, such as the big bang, where physical quantities like curvature and energy density diverge. The resolution of cosmological singularities remains one of the central challenges in theoretical cosmology. Quantum gravity is expected to resolve these singularities, and various approaches, including loop quantum cosmology \cite{ashtekar2006quantum, bojowald2005loop}, generalized uncertainty principles \cite{kempf1995hilbert, scardigli1999generalized}, and polymer quantization \cite{corichi2007polymer}, have provided scenarios in which such divergences are replaced by smooth bounces. In particular, polymer quantization introduces a fundamental length scale via the substitution of momenta with trigonometric functions, leading to modified Hamiltonian dynamics that remain well-defined near Planckian regimes. 

In polymer quantum approach a polymer length scale, $\mu$, which shows the scale of the segments of the granular space enters into the Hamiltonian of the system to deform its functional form into a so-called polymeric Hamiltonian. This means that in a polymeric quantized system in addition of a quantum parameter $\hbar$, which is responsible to canonical quantization of the system, there is also another quantum parameter $\mu$, that labels the granular properties of the underlying space. This approach then opened a new windows for the
theories which are dealing with the quantum gravitational effects in physical systems such as quantum cosmology and black hole physics, see for instance \cite{qc pol} and \cite{pol bl} and the references therein.

To polymerize a dynamical system one usually begins with a classical system described by Hamiltonian $H$. The canonical quantization of such a system transforms its Hamiltonian to an Hermitian operator, which now contains the parameter $\hbar$, in such a way that in the limit $\hbar \rightarrow 0$, the quantum Hamiltonian $H_{\hbar}$ returns to its classical counterpart. By polymerization, the Hamiltonian gets an additional quantum parameter $\mu$, which is rooted in the ideas of granular structure of the space-time. Therefore, by taking the classical limit of the resulting Hamiltonian $H_{\hbar,\mu}$, we arrive at a semi-classical theory in which the parameter $\mu$ is still present. To achieve the initial classical theory, one should once again take the limit $\mu \rightarrow 0$ from this intermediate theory. It is believed that such an effective classical theories $H_{\mu}$,
have enough rich structure to exhibit some important features of the system related to the quantum effects without quantization of the
system. The process by which the theory $H_{\mu}$ is obtained from the classical theory is called {\it classical polymerization}.
A detailed explanation of this process with some of its cosmological applications can be found in \cite{CPR}.

Another line of investigation involves deforming the symplectic structure of phase space, either phenomenologically or as an effective description arising from quantum gravity or non-commutative geometry \cite{ali2009discreteness, vakili2008classical, Barca}. Such deformations can be encoded via modified Poisson brackets, leading to generalized Hamiltonian systems.
In cosmology, while most studies have focused on isotropic models, there are also works have begun exploring the role of polymer effects in anisotropic cosmologies \cite{chiou2007loop}, including Bianchi models. Anisotropic cosmological models, provide a natural testing ground for exploring quantum gravitational effects, since they generalize the isotropic Friedmann--Robertson--Walker (FRW) models and can capture more complex dynamical behaviors near the Planck scale. A broad overview of quantum-gravity inspired modifications of cosmological dynamics can be found in 
\cite{ijgmmp2025review,indian2021anisotropy, ijgmmp2021qg}, while various applications to anisotropic models 
are discussed in \cite{indian2023polymer, ijgmmp2023review}. 
Related approaches combining polymer quantization or deformed structures with cosmological scenarios 
include for example in \cite{ijgmmp2024deformed, indian2019bianchi}.

In this paper, we study the classical dynamics of the Bianchi~I model in the presence of both 
polymer quantization and an exponential deformation of the Poisson structure. We are motivated by the question of how effective, quantum-gravity-inspired modifications of the canonical phase space affect the dynamics of the early Universe. In particular, volume-dependent deformations of the Poisson algebra provide a simple but physically motivated ansatz to model state-dependent discreteness or scale-dependent corrections that may arise in loop-inspired frameworks. Anisotropic cosmologies (such as Bianchi type I) furnish the minimal setting in which the nontrivial effects of such deformations can be probed, since the anisotropy degrees of freedom are absent in the homogeneous and isotropic FRW model. 

The research hypothesis we test in this paper is the following: the combined action of classical polymerization and an exponential, volume-dependent deformation of the Poisson structure will (i) alter the effective evolution rate of the logarithmic volume $\alpha(t)$ (typically slowing its slope relative to the undeformed case) and (ii) modify the anisotropy dynamics in a way that can stabilize the shear degrees of freedom (i.e. render $\beta_\pm(t)$ bounded) for a suitable range of deformation parameters $s_i$. To verify this hypothesis we derive the polymer--deformed effective Hamiltonian, obtain analytic expressions for $\alpha(t)$ and $\beta_\pm(t)$ in a convenient gauge, and support the analytic results with numerical examples illustrating the qualitative changes induced by the deformation.

It is worth noting why we work with the Bianchi~I geometry instead of the spatially flat FRW model. 
The FRW universe contains only a single degree of freedom (the scale factor), and any deformation 
of the Poisson algebra would essentially reduce to a trivial rescaling of this variable. 
By contrast, the Bianchi~I model includes two anisotropy degrees of freedom $\beta_\pm$ in addition 
to the logarithmic volume $\alpha$, providing the minimal setting in which the impact of a 
volume-dependent deformation of the Poisson brackets can be meaningfully tested. 
Physically, this choice is also motivated by the possibility that the early universe contained 
small anisotropies, and by studying their suppression one gains insight into the observed 
near-isotropy of the present cosmos. The isotropic FRW case is recovered as the special limit 
$\beta_\pm \to 0$ of our solutions.

Starting from a general Hamiltonian, we implement the polymerization 
in the canonical momenta and introduce the deformation through 
$\{q_i,p_j\} = \delta_{ij} e^{s_i \alpha}$. 
We derive the corresponding equations of motion and obtain analytic solutions for the volume variable 
$\alpha(t)$, and the anisotropy parameters $\beta_\pm(t)$. 
Our analysis focuses on the contracting branch, where we find that the exponential deformation with 
appropriate $s_i$ values does not remove the initial singularity but leads to a slower evolution 
of the volume compared to the undeformed model and, remarkably, to the stabilization of anisotropy 
variables, which remain bounded throughout the evolution. 
These results are supported by numerical plots, which clearly illustrate the qualitative differences 
between the standard polymerized Bianchi~I cosmology and its exponentially deformed counterpart.

\section{Bianchi type I}
Let us make a quick review of some of the important results in the Bianchi type I model and obtain its Hamiltonian in the ADM decomposition, for more details see \cite{Gron}. The Bianchi models are the most general homogeneous cosmological solutions of the Einstein field equations which admit a three-dimensional isometry group, i.e. their spatially
homogeneous sections are invariant under the action of a three-dimensional Lie group. In the Misner notation \cite{Misner}, the metric of the Bianchi models can be written as

\begin{equation}
ds^2 = -N^2(t) dt^2 + h_{ab}(t)\,\omega^a \otimes \omega^b,
\end{equation}
where \(N(t)\) is the lapse function, \(h_{ab}(t)\) is the spatial metric depending only on time, and \(\omega^a\) are the left-invariant 1-forms of the corresponding
isometry group on the spatial manifold satisfying the Maurer-Cartan equations

\begin{equation}
d\omega^{a} = -\frac12 C^{a}{}_{bc}\,\omega^{b} \wedge \omega^{c},
\end{equation}
where \(C^{a}{}_{bc}\) are the structure constants of the corresponding three-dimensional Lie algebra. In Bianchi type I all \(C^{a}{}_{bc} = 0\). A convenient parameterization of the spatial metric \(h_{ab}\) uses the Misner variables

\begin{equation}
h_{ab} = e^{2\alpha(t)}\,e^{2\beta_{ab}(t)}, \qquad \mathrm{tr}\,\beta=0,
\end{equation}which transforms the Hamiltonian of the dynamical system corresponds
to the Bianchi model to a more manageable form. The traceless anisotropy tensor \(\beta_{ab}\) can be described by two independent parameters \(\beta_{+}(t)\) and \(\beta_{-}(t)\), leading to the diagonal form

\begin{align}
\beta_{ab} = \mathrm{diag}\big(\beta_+ + \sqrt{3}\,\beta_-,\,\beta_+ - \sqrt{3}\,\beta_-,\, -2\beta_+ \big).
\end{align}
Here, \(\alpha\) encodes the logarithmic volume of the universe, while \(\beta_{\pm}\) describe the anisotropic shear. In terms of these variables the line element becomes

\begin{equation}
ds^2 = -N^2(t) dt^2 + e^{2\alpha(t)}\,[e^{2\beta(t)}]_{ab}\,\omega^a\omega^b.
\end{equation}
The Einstein-Hilbert action is given by (we work in units where
$c=16\pi G=1$)

\begin{equation}\label{D}
{\cal S}=\int d^4 x\sqrt{-g}{\cal R},\end{equation}
where $g$ is the determinant and ${\cal R}$ is the
scalar curvature of the space-time metric. In terms of the ADM variables,
action (\ref{D}) can be written as \cite{Vakili}

\begin{equation}\label{E}
{\cal S}=\int dt d^3x{\cal L}=\int dt d^3x
N\sqrt{h}\left(K_{ab}K^{ab}-K^2+R\right),\end{equation}
where $K_{ab}$ are the components of extrinsic curvature (second
fundamental form) which represent how much the spatial space
$h_{ab}$ is curved in the way it sits in the space-time manifold.
Also, $h$ and $R$ are the determinant and scalar curvature of the
spatial geometry $h_{ab}$ respectively, and $K$ represents the
trace of $K_{ab}$. The extrinsic curvature is given by

\begin{equation}\label{F}
K_{ab}=\frac{1}{2N}\left(N_{a|b}+N_{b|a}-\frac{\partial
h_{ab}}{\partial t}\right),\end{equation}where $N_{a|b}$
represents the covariant derivative with respect to $h_{ab}$.
Using the form of the metric and definition of $\beta_{ab}$ we obtain the non-vanishing
components of the extrinsic curvature and its trace as follows

\begin{eqnarray}\label{G}
K_{11}&=&-\frac{1}{N}(\dot{\alpha}+\dot{\beta_{+}}+\sqrt{3}\dot{\beta_{-}})e^{2(\alpha+\beta_{+}+\sqrt{3}\beta_{-})},\nonumber \\
K_{22}&=&-\frac{1}{N}(\dot{\alpha}+\dot{\beta_{+}}-\sqrt{3}\dot{\beta_{-}})e^{2(\alpha+\beta_{+}-\sqrt{3}\beta_{-})},\nonumber\\
K_{33}&=& -\frac{1}{N}(\dot{\alpha}-2\dot{\beta_{+}})e^{2(\alpha-2\beta_{+})},\nonumber\\
K&=&-3\frac{\dot{\alpha}}{N}, \nonumber
\end{eqnarray}
where a dot represents differentiation with respect to $t$. The
scalar curvature $R$ of a  spatial hypersurface is a function of
$\beta_{+}$ and $\beta_{-}$ and can be write in terms of the structure constants
as 
\begin{equation}\label{H}
R=C^a_{bc}C^l_{mn}h_{a}h^{cm}h^{bn}+2C^a_{bc}C^c_{la}h^{bl}.
\end{equation}
The Lagrangian for the Bianchi models may now be written
by substituting the above results into action (\ref{E}), giving
\begin{equation}\label{I}
{\cal
L}=\frac{6e^{3\alpha}}{N}\left(-\dot{\alpha}^2+\dot{\beta_{+}}^2+\dot{\beta_{-}}^2\right)+Ne^{3\alpha}R.
\end{equation}
The momenta conjugate to the dynamical variables are given by
\begin{equation}\label{J}
p_{\alpha}=\frac{\partial {\cal L}}{\partial
\dot{\alpha}}=-\frac{12}{N}\dot{\alpha}e^{3\alpha},\hspace{.5cm}p_{+}=\frac{\partial
{\cal L}}{\partial
\dot{\beta_{+}}}=\frac{12}{N}\dot{\beta_{+}}e^{3\alpha},\hspace{.5cm}p_{-}=\frac{\partial
{\cal L}}{\partial \dot{\beta_{-}}}=\frac{12}{N}\dot{\beta_{-}}e^{3\alpha},
\end{equation}
leading to the following Hamiltonian

\begin{equation}\label{K}
H=N{\cal H}=\frac{1}{24}Ne^{-3\alpha}\left(-p_{\alpha}^2+p_{+}^2+p_{-}^2\right)-Ne^{3\alpha}R.
\end{equation}
The preliminary set-up for writing the dynamical equations is now complete. The cosmological dynamics of the Bianchi models are studied in many works, see for example \cite{works} and the references therein. In what follows, we consider the Bianchi I model for which $R=0$, and will study these equations in the framework of classical polymerization combined with a deformed Poisson algebra.

\section{A Brief Review of Classical Polymerization}
In Schr\"{o}dinger picture of quantum mechanics, the coordinates and momentum representations are equivalent and may be easily converted to each other by a Fourier transformation.
However, in the presence of the quantum gravitational effects the space-time may take a discrete structure so that such a well-defined representations are no longer applicable. As an alternative, polymer quantization provides a suitable framework for studying these situations \cite{corichi2007polymer, ashtekar2003mathematical}. The Hilbert space of this
representation of quantum mechanics is ${\mathcal H}_{\rm poly}=L^2(R_{_d},d\mu_{_d})$, where $d\mu_{_d}$ is the Haar measure, and $R_{_d}$ denotes the real discrete line whose segments are
labeled by an extra dimension-full parameter $\mu$ such that the standard Schr\"{o}dinger picture will be recovered in the continuum limit $\mu\rightarrow\ 0$. This means that by a classical limit $\hbar\rightarrow\,0$, the polymer quantum mechanics tends to an effective $\mu$-dependent classical theory which is somehow different from the classical theory from which we have started. Such an effective theory may also be obtained directly from the standard
classical theory, without referring to the polymer quantization, by using of the Weyl operator \cite{CPR}. The process is known as {\it polymerization} with which we will deal in the rest of this paper.

According to the mentioned above form of the Hilbert space of the polymer representation of quantum mechanics, the position space (with coordinate $q$) has a discrete structure with discreteness
parameter $\mu$. Therefore, the associated momentum operator $\hat{p}$, which is the generator of the displacement, does not exist \cite{ashtekar2003mathematical, banerjee2007discreteness}. However, the Weyl exponential operator (shift operator) correspond to the discrete translation along $q$ is well defined and effectively plays the role of momentum associated to $q$
\cite{corichi2007polymer}. This allows us to utilize the Weyl operator to find an effective momentum in the semiclassical regime. So, consider a state $f(q)$, its derivative with respect to the discrete position $q$ may be approximated by means of the Weyl operator as \cite{CPR}

\begin{eqnarray}\label{FWD}
\partial_{q}f(q)\approx\frac{1}{2\mu}[f(q+\mu)-f(q-
\mu)]\hspace{2cm}\nonumber\\=\frac{1}{2\mu}\Big(
\widehat{e^{ip\mu}}-\widehat{e^{-ip\mu}}\Big)\,f(q)=
\frac{i}{\mu}\widehat{\sin(\mu p)}\,f(q),
\end{eqnarray}
and similarly the second derivative approximation will be
\begin{eqnarray}\label{SWD}
\partial_{q}^2f(q)\approx\frac{1}{\mu^2}[f(q+\mu)-2
f(q)+f(q-\mu)]\hspace{1cm}\nonumber\\=\frac{2}{\mu^2}
(\widehat{\cos(\mu p)}-1)\,f(q).\hspace{2cm}
\end{eqnarray}
Having the above approximations at hand, we define the polymerization process for the finite values of the parameter
$\mu$ as

\begin{eqnarray}\label{Polymerization}
\hat{p}\rightarrow\,\frac{1}{\mu}\widehat{\sin(\mu p)},
\hspace{1cm}\hat{p}^2\rightarrow\,\frac{2}{\mu^2}(1-
\widehat{\cos(\mu p)}).
\end{eqnarray}This replacements suggest the idea that a classical theory may be obtained via this process, but now without any attribution to the Weyl operator. This is what which is dubbed
usually as {\it classical Polymerization} in literature
\cite{corichi2007polymer, CPR}:

\begin{eqnarray}\label{PT}
q\rightarrow q,\hspace{1.5cm}p\rightarrow\frac{ \sin(\mu
p)}{\mu},\hspace{5mm}p^2\rightarrow
\frac{2}{\mu^2}\left[1-\cos(\mu p)\right],
\end{eqnarray}where now $(q,p)$ are a pair of classical phase space variables. Hence, by applying the transformation (\ref{PT}) to the Hamiltonian of a classical system we get its classical polymerized counterpart. A glance at (\ref{PT}) shows that the momentum is periodic and varies in a bounded interval as $p\in[-\frac{\pi}{\mu},+\frac{\pi}{ \mu})$. In the limit
$\mu\rightarrow\,0$, one recovers the usual range for the canonical momentum $p\in(-\infty,+\infty)$. Therefore, the polymerized momentum is compactified and topology of the momentum
sector of the phase space is $S^1$ rather than the usual $\mathbb{R}$ \cite{NaturalCutoff}. Our set-up to explain the classical polymerization of a dynamical system is now complete. This procedure has been applied to various cosmological models as an effective way to incorporate some aspects of quantum gravitational effects without relying on the full quantum formalism. In particular, when applied to homogeneous cosmologies, the resulting effective dynamics often exhibit a bounce replacing the initial big-bang singularity \cite{bojowald2001absence}.

In the present work, we adopt this classical polymerization scheme and apply it to the Hamiltonian formulation of Bianchi I described above. Furthermore, we generalize the phase space structure by introducing a deformation of the canonical Poisson bracket

\begin{equation}
\{q, p\} = f(q,p),
\end{equation}
where $f(q,p)$ is a function encoding the deformation of the symplectic structure. This so-called Deformed Canonical Relation (DCR) has been previously considered in the context of black hole thermodynamics and effective quantum mechanics \cite{vacaru2008deformation}. When applied to cosmology, it provides a rich framework in which the bounce and the resolution of singularities can be analyzed in a more general setting.

In the following sections, we implement this formalism in the case of Bianchi I cosmology and study the resulting effective dynamics. Our methodology proceeds in the following steps. 
We start from the classical Hamiltonian formulation of the Bianchi~I cosmological model. 
Classical polymerization is then implemented at the effective level by replacing the canonical momenta with their 
polymerized counterparts, introducing characteristic scales $\mu_\alpha$ and $\mu_\pm$. 
In parallel, we deform the Poisson brackets according to $\{q_i,p_j\}=\delta_{ij}\, e^{s_i\alpha},$
which incorporates a volume-dependent exponential factor governed by deformation parameters $s_i$. 
From this deformed phase-space structure we derive the effective equations of motion for the volume variable $\alpha(t)$ 
and the anisotropy parameters $\beta_\pm(t)$. These equations admit analytic solutions and the qualitative behavior is illustrated 
with numerical plots. Our methodology therefore combines analytic and numerical tools to highlight the distinctive impact of the 
volume-dependent deformation on the polymerized Bianchi~I dynamics.

\section{Polymerization, DCR and Equations of Motion}
As explained above the method of polymerization is based on the modification of the Hamiltonian to get a deformed Hamiltonian $H_{\mu}$ where $\mu$ is the deformation
parameter. For our system this method will be done by applying the transformation (\ref{PT}) on the Hamiltonian (\ref{K}). So, by means of the transformation

\begin{equation}
p_i \ \longrightarrow\ {\cal P}_{i}(p_i)=\frac{\sin(\mu_i p_i)}{\mu_i}, 
\quad i \in \{\alpha, +, -\},
\label{eq:polymerization}
\end{equation}
the polymerized Hamiltonian for the Bianchi type I model takes the form

\begin{equation}
H=N\mathcal{H}_{\text{poly}} = N e^{-3\alpha} \left[ 
 -\frac{\sin^2(\mu_\alpha p_\alpha)}{\mu_\alpha^2}
 + \frac{\sin^2(\mu_+ p_+)}{\mu_+^2}
 + \frac{\sin^2(\mu_- p_-)}{\mu_-^2}\right],
\label{eq:H_poly}
\end{equation}
where \(\mu_i\) are the polymerization scales associated with each canonical variable. 

As we mentioned earlier, by this one-parameter
$\mu$-dependent classical theory, we expect to address the quantum features of the system without a direct reference to the quantum mechanics. Indeed, here instead of first dealing with the
quantum pictures based on the quantum Hamiltonian operator, one modifies the classical Hamiltonian according to the transformation (\ref{PT}) and then deals with classical dynamics of the system with this deformed Hamiltonian. In the resulting classical system the discreteness parameter $\mu$ plays an essential role since its supports the idea that the $\mu$-correction to the classical
theory is a signal from quantum gravity. Under these conditions the Hamiltonian equations of motion for the above Hamiltonian are $\dot{q}_i=\{q_i,\,N\mathcal{H}_{\rm poly}\}$. To compute the Poisson brackets, we allow a deformation of the canonical Poisson brackets of the (simple) diagonal form

\begin{equation}\label{deformed_poisson}
\{q^i,p_j\} = \delta^{i}_{j}\,g_i(q),\qquad
\{q_i,q_j\}=0,\qquad \{p_i,p_j\}=0,
\end{equation}
where \(g_i(q)\) are nonzero scalar functions of the configuration variables \(q=(\alpha,\beta_+,\beta_-)\). The usual canonical bracket is recovered for \(g_i\equiv 1\). In this paper we adopt a volume-dependent deformation of the Poisson algebra

\begin{equation}\label{deform}
\{q^i,p_j\}=\delta^i{}_j\,g_i(\alpha),\qquad g_i(\alpha)=e^{s_i\alpha},
\end{equation}
with dimensionless deformation parameters $s_i$ (the canonical algebra is recovered for $s_i=0$). This equation generalizes the canonical relation $\{q_i,p_j\}=\delta_{ij}$ by allowing an explicit 
dependence on the logarithmic volume $\alpha$. The exponential form is chosen as the simplest ansatz encoding scale-dependent corrections, with $s_i$ denoting the deformation parameters.

With lapse function \(N(t)\) the dynamics is generated by \(H=N\mathcal{H}_{\rm poly}\). Using \eqref{deformed_poisson}, the equations of motion take the compact form
\begin{align}
\dot{q}_i &= \{q_i,\,N\mathcal{H}_{\rm poly}\}
= N\,\{q_i,p_k\}\,\frac{\partial\mathcal{H}_{\rm poly}}{\partial p_k}
= N\,g_i(q)\,\frac{\partial\mathcal{H}_{\rm poly}}{\partial p_i}, \label{qdot_general}\\[4pt]
\dot{p}_i &= \{p_i,\,N\mathcal{H}_{\rm poly}\}
= -N\,\{q_k,p_i\}\,\frac{\partial\mathcal{H}_{\rm poly}}{\partial q_k}
= -N\,g_i(q)\,\frac{\partial\mathcal{H}_{\rm poly}}{\partial q_i}. \label{pdot_general}
\end{align}
Since \(\mathcal{H}_{\rm poly}\) depends on \(p_i\) only through \(\mathcal{P}_i(p_i)\), one useful derivative identity is
\begin{equation}
\frac{\partial\mathcal{H}_{\rm poly}}{\partial p_i}
= \frac{d\mathcal{P}_i}{dp_i}\,\frac{\partial\mathcal{H}_{\rm poly}}{\partial \mathcal{P}_i}
= \cos(\mu_i p_i)\,\frac{\partial\mathcal{H}_{\rm poly}}{\partial \mathcal{P}_i},
\end{equation}
with \(\dfrac{d\mathcal{P}_i}{dp_i}=\cos(\mu_i p_i)\). The physical trajectories must also satisfy the polymerized Hamiltonian constraint

\begin{equation}
\mathcal{H}_{\rm poly}(q,p)=0.
\end{equation}
Since the derivatives of the Hamiltonian with respect to the momenta are

\begin{align}
\frac{\partial\mathcal{H}_{\rm poly}}{\partial p_\alpha}
&= N e^{-3\alpha}\left[-\frac{\sin(2\mu_\alpha p_\alpha)}{\mu_\alpha}\right],\\[4pt]
\frac{\partial\mathcal{H}_{\rm poly}}{\partial p_\pm}
&= N e^{-3\alpha}\left[ \frac{\sin(2\mu_\pm p_\pm)}{\mu_\pm}\right],
\end{align}
using $\dot q^i = g_i(\alpha)\,\partial\mathcal{H}_{\rm poly}/\partial p_i$, we obtain

\begin{align}
\dot{\alpha} &= -\,N\,e^{(s_\alpha-3)\alpha}\,\frac{\sin(2\mu_\alpha p_\alpha)}{\mu_\alpha},\label{alpha_dot}\\[4pt]
\dot{\beta}_\pm &= \ N\,e^{(s_\pm-3)\alpha}\,\frac{\sin(2\mu_\pm p_\pm)}{\mu_\pm}.\label{beta_dot}
\end{align}
The derivatives with respect to configuration variables reduce to the $\alpha$-dependence through the overall factor $e^{-3\alpha}$. Denoting

\begin{equation}
\mathcal{B} \equiv -\frac{\sin^2(\mu_\alpha p_\alpha)}{\mu_\alpha^2}
 + \frac{\sin^2(\mu_+ p_+)}{\mu_+^2}
 + \frac{\sin^2(\mu_- p_-)}{\mu_-^2},
\end{equation}
one finds

\begin{equation}
\frac{\partial\mathcal{H}_{\rm poly}}{\partial \alpha} = -3N e^{-3\alpha}\,\mathcal{B}.
\end{equation}
Hence the momentum equations are

\begin{align}
\dot{p}_\alpha &= -\,g_\alpha(\alpha)\,\frac{\partial\mathcal{H}_{\rm poly}}{\partial \alpha}
= 3N e^{(s_\alpha-3)\alpha}\,\mathcal{B},\label{palpha_dot}\\[4pt]
\dot{p}_\pm &= -\,g_\pm(\alpha)\,\frac{\partial\mathcal{H}_{\rm poly}}{\partial \beta_\pm}
= 0,\label{p_pm_dot}
\end{align}
which means that $p_+$ and $p_-$ are constants of motion in the Bianchi I model.

The Hamiltonian constraint $\mathcal{H}_{\rm poly}=0$ implies $\mathcal{B}=0$, or explicitly

\begin{equation}\label{constraint_relation}
\frac{\sin^2(\mu_\alpha p_\alpha)}{\mu_\alpha^2}
=\frac{\sin^2(\mu_+ p_+)}{\mu_+^2}
+\frac{\sin^2(\mu_- p_-)}{\mu_-^2}.
\end{equation}
Since $p_\pm$ are constant, \eqref{constraint_relation} determines $p_\alpha$ (up to discrete branches)
as a function of the conserved anisotropy momenta. In the classical (continuum) limit $\mu_i\to 0$, we have $\sin(\mu_i p_i)/\mu_i \to p_i$, and
\eqref{constraint_relation} reduces to the usual Bianchi~I relation
\begin{equation}
-p_\alpha^2 + p_+^2 + p_-^2 = 0.
\end{equation}
Before addressing the solution of the equations of motion, it is important to emphasize a few points:

\begin{itemize}
\item The constants $p_\pm$ parametrize the anisotropic conserved momenta: different choices label different solutions; $p_\alpha$ is then fixed by the constraint.
\item The deformation factors $s_i$ modify the relation between proper time (or chosen slicing) and phase-space flow: e.g. $\dot\alpha$ and $\dot\beta_\pm$ acquire factors $e^{(s_i-3)\alpha}$
    which can accelerate or slow down the approach to singular behavior depending on the signs of $s_i$.
\item For explicit solutions one may (i) choose a gauge for the lapse $N(t)$ (e.g. $N=1$ or $N=e^{3\alpha}$), (ii) solve the constraint for $p_\alpha$, and (iii) integrate \eqref{alpha_dot} and
    \eqref{beta_dot}. The presence of $\sin(2\mu_i p_i)$ may lead to bounded oscillatory behavior for the configuration variables compared to the unbounded classical trajectories.
\end{itemize}

\section{Analytic Solutions in a Convenient Gauge}
The above dynamical equations of motion in view of the concerning issue of time, has been of course under-determined. Before trying to solve
these equations we must decide on a choice of time in the theory. The under-determinacy problem at the classical level may be removed by using the gauge freedom via fixing the
gauge. Here, we choose the harmonic time lapse gauge $N(t)=e^{3\alpha(t)}$, so that the polymerized Hamiltonian reduces (up to the overall factor) to the constraint

\begin{equation}
\mathcal{B} \equiv
 -\frac{\sin^2(\mu_\alpha p_\alpha)}{\mu_\alpha^2}
 + \frac{\sin^2(\mu_+ p_+)}{\mu_+^2}
 + \frac{\sin^2(\mu_- p_-)}{\mu_-^2} = 0.
\end{equation}The ebove equation follows from the polymerized Hamiltonian constraint. 
In its derivation we have fixed the lapse $N=1$ and selected the sign of the Hamiltonian constraint 
corresponding to the contracting branch. This relation thus represents the effective evolution of 
the anisotropy variables $\beta_\pm$ in the polymer--deformed Bianchi~I dynamics. Therefore, the equations of motion obtained earlier become with this gauge

\begin{align}
\dot{\alpha} &= -\,e^{s_\alpha\alpha}\,\frac{\sin(2\mu_\alpha p_\alpha)}{\mu_\alpha}, \label{eq:alpha_dot_gauge}\\[4pt]
\dot{\beta}_\pm &= \ e^{s_\pm\alpha}\,\frac{\sin(2\mu_\pm p_\pm)}{\mu_\pm},\label{eq:beta_dot_gauge}\\[4pt]
\dot{p}_\alpha &= 3 e^{s_\alpha\alpha}\,\mathcal{B},\qquad
\dot{p}_\pm = 0. \label{eq:p_constants}
\end{align}
From \(\dot p_\pm=0\) we have

\begin{equation}
p_+ = P_+,\qquad p_- = P_- \quad(\text{constants}).
\end{equation}
The constraint \(\mathcal{B}=0\) then fixes \(p_\alpha\) algebraically

\begin{equation}\label{eq:palph_const}
\frac{\sin^2(\mu_\alpha p_\alpha)}{\mu_\alpha^2}
= C,\qquad
C \equiv \frac{\sin^2(\mu_+ P_+)}{\mu_+^2} + \frac{\sin^2(\mu_- P_-)}{\mu_-^2}.
\end{equation}
Consistency requires \(0\le \mu_\alpha^2 C \le 1\), otherwise no real solution for \(p_\alpha\) exists. Choosing a branch for the arcsine

\begin{equation}
\sin(\mu_\alpha p_\alpha) = \sigma\,\mu_\alpha\sqrt{C},\qquad \sigma=\pm 1,
\end{equation}
we obtain discrete branches for \(p_\alpha\)

\begin{equation}
p_\alpha = \frac{1}{\mu_\alpha}\left[(-1)^k \arcsin(\sigma\,\mu_\alpha\sqrt{C}) + k\pi\right],\qquad k\in\mathbb{Z}.
\end{equation}
For any fixed branch, \(p_\alpha\) is a constant of motion.

With constant \(p_\alpha\), the factor \(\sin(2\mu_\alpha p_\alpha)\) is constant. Define

\begin{equation}
K \equiv -\frac{\sin(2\mu_\alpha p_\alpha)}{\mu_\alpha}.
\end{equation}
Using \(\sin(\mu_\alpha p_\alpha)=\sigma\,\mu_\alpha\sqrt{C}\), one gets

\begin{equation}
K = -2\sigma\sqrt{C}\,\sqrt{1-\mu_\alpha^2 C},
\end{equation}
which is a real constant provided the constraint \( \mu_\alpha^2 C\le 1\) holds. Thus \eqref{eq:alpha_dot_gauge} becomes a separable differential equation as
 
\begin{equation}
\dot\alpha = K\, e^{s_\alpha \alpha}.
\end{equation}
If \(s_\alpha \neq 0\), This equation admits the solution

\begin{equation}\label{eq:alpha_solution_nonzero_s}
\alpha(t) = -\frac{1}{s_\alpha}\ln\big[- s_\alpha (K t + C_0)\big],
\end{equation}where \(C_0\) is an integration constant which sets the time origin. This solution is real on time intervals where the argument of the log is positive

\begin{equation}
- s_\alpha (K t + C_0) > 0.
\end{equation}
If \(s_\alpha = 0\), then \(\dot\alpha = K\) and

\begin{equation}\label{eq:alpha_solution_s_zero}
\alpha(t) = K t + \alpha_0,
\end{equation}
with \(\alpha_0\) an integration constant. Note that different branches of the arcsine in \eqref{eq:palph_const} correspond to discrete families of solutions labeled by integers \(k\). These may lead to different signs for \(K\) and hence expanding vs contracting behaviors.

To obtain $\beta_{\pm}(t)$, recall

\begin{equation}
\dot\beta_\pm = D_\pm\, e^{s_\pm \alpha(t)},\qquad
D_\pm \equiv \frac{\sin(2\mu_\pm P_\pm)}{\mu_\pm} \quad(\text{constants}).
\end{equation}

Again if \(s_\alpha\neq 0\), use \eqref{eq:alpha_solution_nonzero_s} to express \(e^{s_\pm\alpha}\) as

\begin{equation}
e^{s_\pm\alpha(t)} = \big[- s_\alpha (K t + C_0)\big]^{-s_\pm/s_\alpha},
\end{equation}
yields

\begin{equation}
\dot\beta_\pm = D_\pm\big[- s_\alpha (K t + C_0)\big]^{-s_\pm/s_\alpha}.
\end{equation}
Integrating for \(s_\pm/s_\alpha \neq 1\), results

\begin{equation}\label{eq:beta_solution_general}
\beta_\pm(t) = \beta_{\pm,0}
+ \frac{D_\pm}{K}\,\frac{\big[- s_\alpha (K t + C_0)\big]^{1 - s_\pm/s_\alpha}
- \big[- s_\alpha (K t_0 + C_0)\big]^{1 - s_\pm/s_\alpha}}{1 - s_\pm/s_\alpha},
\end{equation}
where \(\beta_{\pm,0}=\beta_\pm(t_0)\), while for \(s_\pm = s_\alpha\), the integral gives a logarithm function 

\begin{equation}\label{eq:beta_solution_log}
\beta_\pm(t) = \beta_{\pm,0} + \frac{D_\pm}{K}\,\ln\frac{K t + C_0}{K t_0 + C_0}.
\end{equation}
In the case where \(s_\alpha=0\), we have \(\alpha(t)=K t+\alpha_0\), and thus

\begin{equation}
\dot\beta_\pm = D_\pm\, e^{s_\pm (K t+\alpha_0)}.
\end{equation}
Therefore

\begin{equation}\label{eq:beta_solution_salpha_zero}
\beta_\pm(t) = \beta_{\pm,0} + \frac{D_\pm}{s_\pm K}\Big[ e^{s_\pm (K t+\alpha_0)} - e^{s_\pm (K t_0+\alpha_0)}\Big],
\end{equation}if $s_\pm\neq 0$, and

\begin{equation}
\beta_\pm(t)=\beta_{\pm,0}+D_\pm (t-t_0),
\end{equation}
if \(s_\pm=0\). 

\begin{figure}[ht]
\centering
\includegraphics[width=0.45\textwidth]{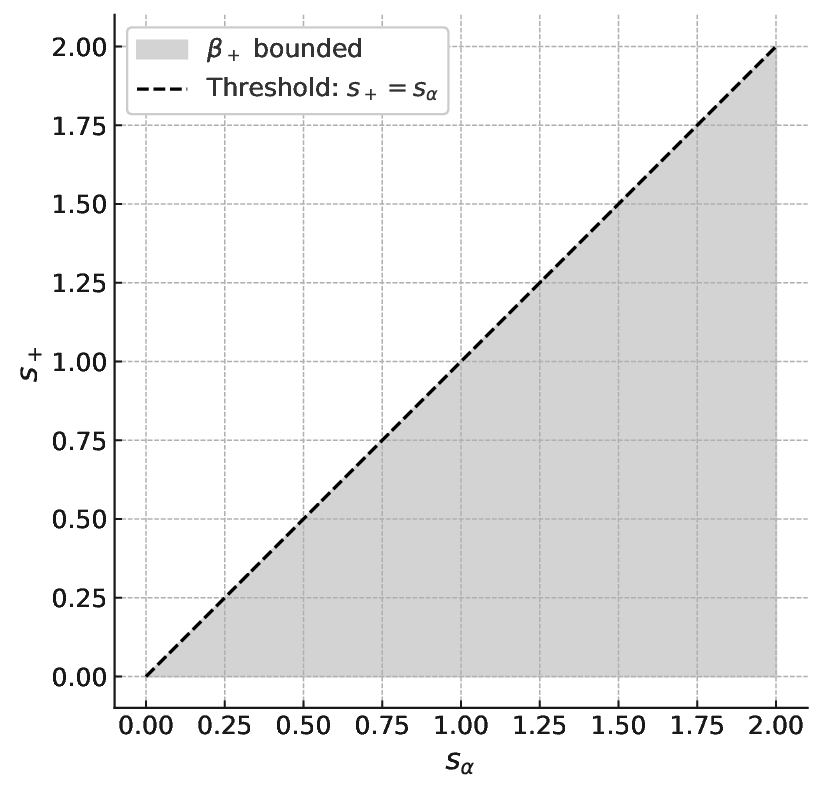}
\caption{Threshold curve in the $(s_\alpha, s_+)$ parameter space. 
The shaded region indicates the set of initial conditions leading 
to bounded anisotropy $\beta_+(t)$ throughout the evolution, while the 
dashed line marks the threshold $s_+ = s_\alpha$.}
\label{fig:threshold_region}
\end{figure}

In this step let us take a look at the analytic threshold for bounded anisotropies. The asymptotic behavior of $\beta_\pm(t)$ near the endpoint where
$A(t)\equiv - s_\alpha (K t + C_0)\to 0^+$, is governed by the exponent
$1-s_\pm/s_\alpha$:
\begin{itemize}
\item If $s_\pm < s_\alpha$, then $1 - s_\pm/s_\alpha>0$ and $\beta_\pm$ tends to a finite limit.
\item If $s_\pm = s_\alpha$, then $\beta_\pm$ grows logarithmically (marginal case).
\item If $s_\pm > s_\alpha$, then $\beta_\pm$ diverges as $A(t)^{\,1 - s_\pm/s_\alpha}\to\infty$.
\end{itemize}
Hence a simple analytic threshold for bounded anisotropies is

\begin{equation}
s_\pm < s_\alpha,
\end{equation}
which defines a critical straight line $s_\pm=s_\alpha$ in the deformation-parameter plane.
Note that the polymer scales $\mu_i$, enter multiplicatively via $D_\pm/K$, and the consistency condition
$\mu_\alpha^2 C\le 1$; these control the amplitude of the response and the domain of validity of the chosen branch
for $p_\alpha$. In particular, if $K\to 0$ (e.g. $\mu_\alpha^2 C\to 1$ or a chosen arcsin branch yields small $\sin(2\mu_\alpha p_\alpha)$),
the prefactor $D_\pm/K$ becomes large and transient effects may be amplified even when $s_\pm<s_\alpha$. The asymptotic analysis reveals the existence of a threshold in the $(s_\alpha,s_+)$ parameter space, separating regimes of bounded and unbounded anisotropy. Figure~\ref{fig:threshold_region} illustrates 
this boundary, highlighting the region where $\beta_+(t)$ remains finite throughout the evolution.

\begin{figure}[ht]
\centering
\includegraphics[width=0.7\textwidth]{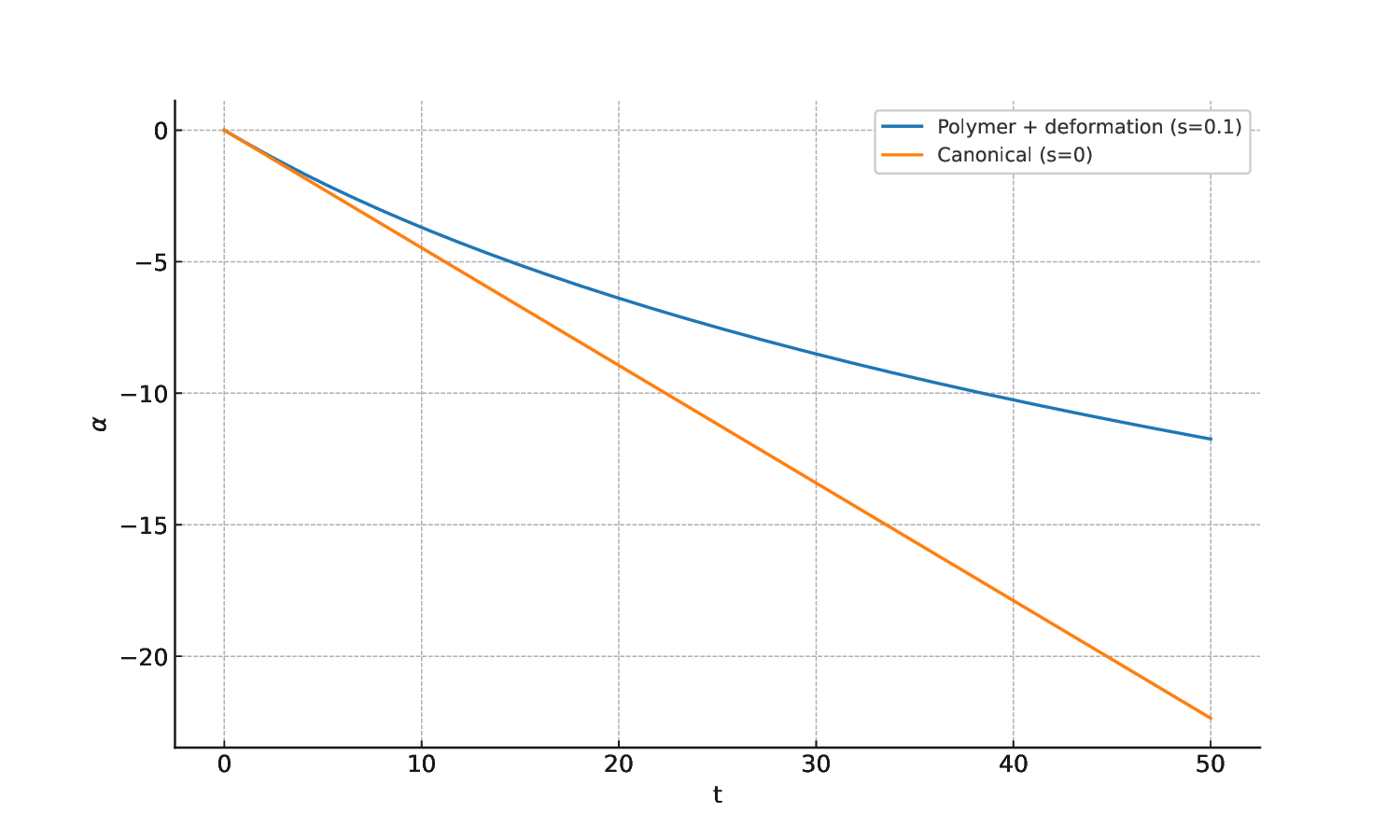}
\caption{Time evolution of $\alpha(t)$ in the polymer-deformed and undeformed Bianchi I models. 
Parameters: $(\mu_\alpha,\mu_+,\mu_-)=(0.1,0.1,0.1)$, $(P_+,P_-)=(0.2,0.1)$, $(s_\alpha,s_+,s_-)=(0.1,0,0)$; 
gauge $N=e^{3\alpha}$. Initial data: $\alpha(0)=0$, $\beta_{\pm}(0)=0$, $t_0=0$. 
Branch choice: principal $\sigma=+1$ for $p_\alpha$ (arcsin branch $k=0$). 
The polymer-deformed curve is generated from Eq.~(47) with $C_0=-1/s_\alpha$ and $K=-\sin(2\mu_\alpha p_\alpha)/\mu_\alpha$, 
on the time domain where $-s_\alpha(Kt+C_0)>0$. The undeformed curve corresponds to $s_\alpha=0$ with the same remaining parameters. The polymer deformation induces 
a slower expansion rate and mild oscillations compared to the standard case.}
\label{fig:alpha_compare}
\end{figure}

\begin{figure}[ht]
\centering
\includegraphics[width=0.6\textwidth]{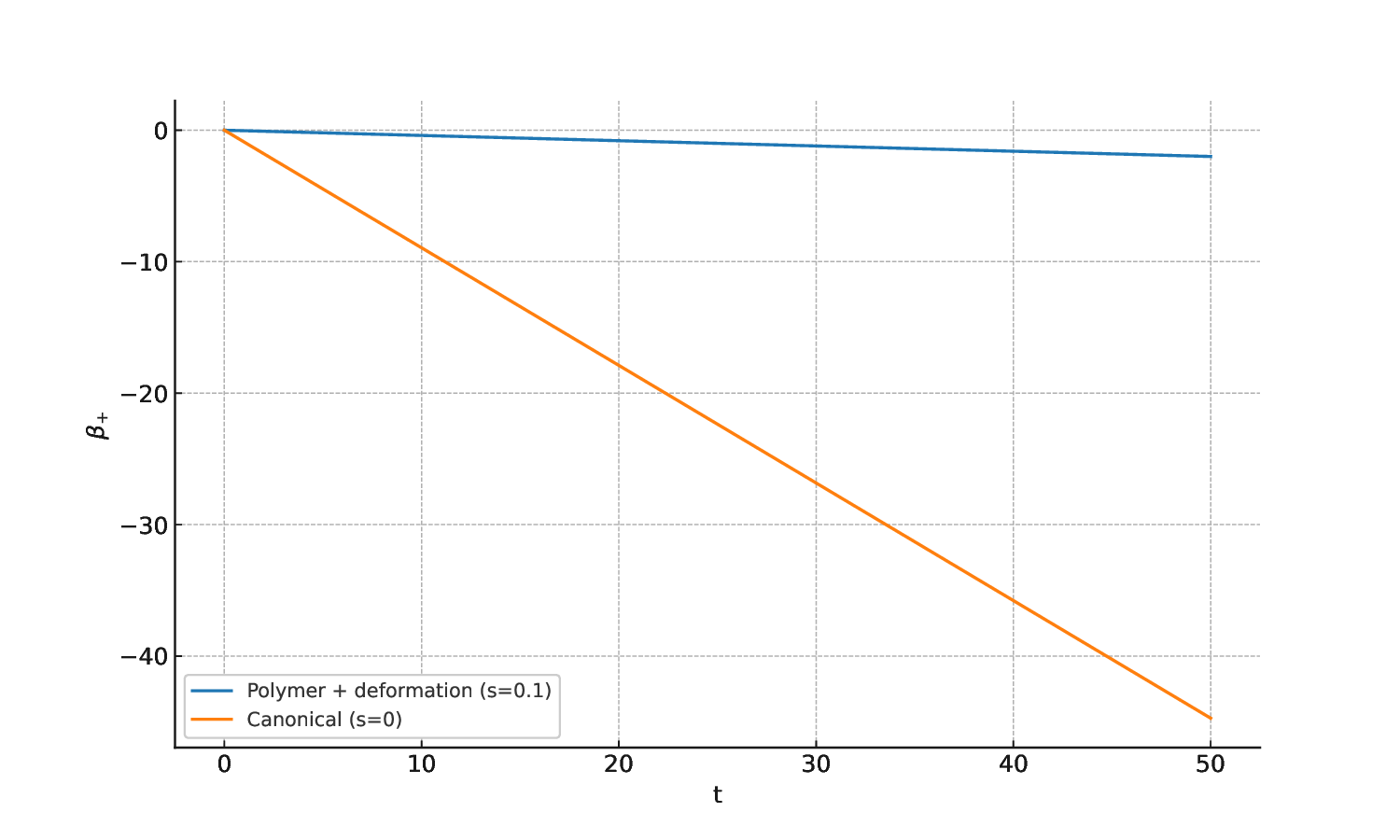}\hspace{25mm}\includegraphics[width=0.6\textwidth]{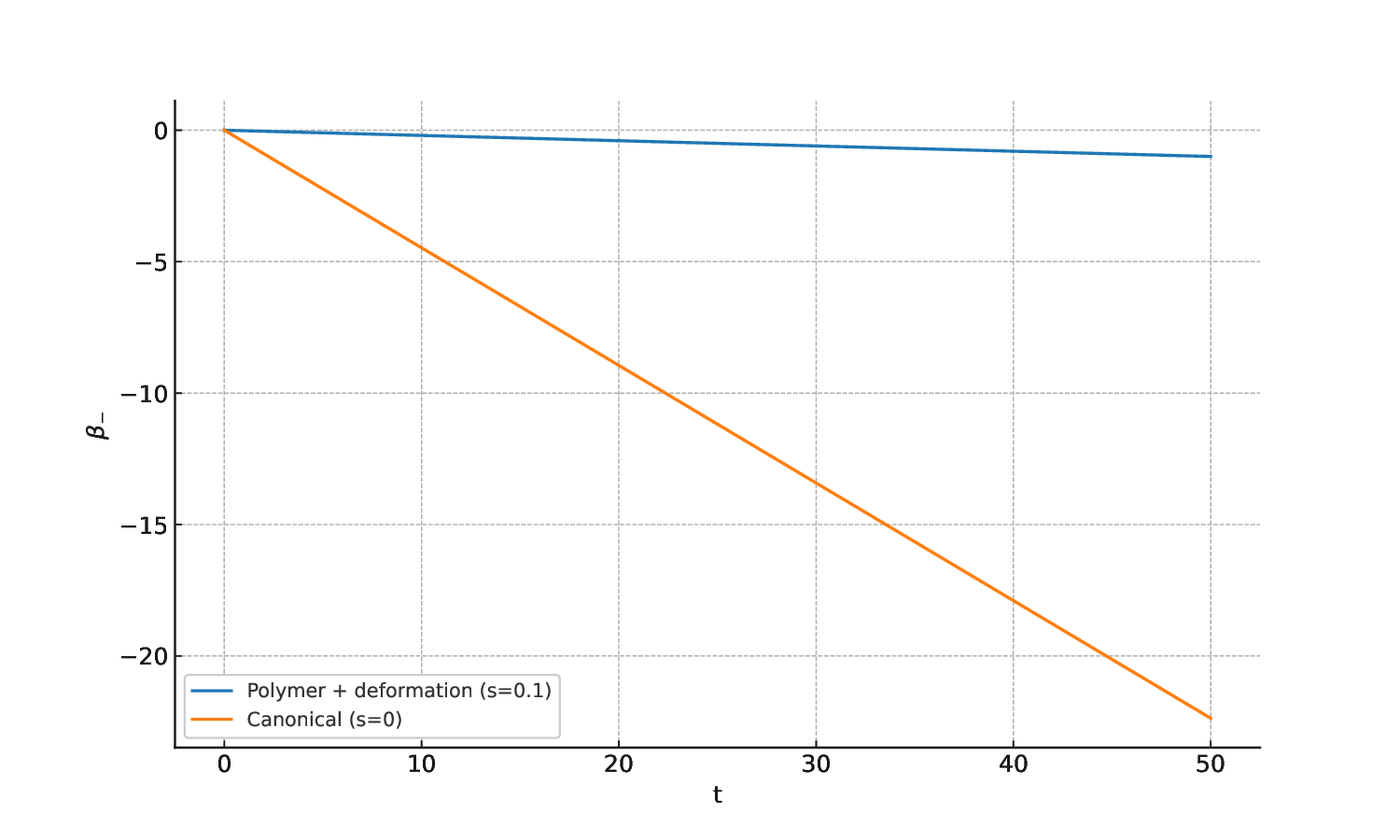}
\caption{Time evolution of $\beta_{\pm}(t)$ in the polymer-deformed and undeformed models. 
Same parameters, gauge and initial data as in Fig.~\ref{fig:alpha_compare}. 
Polymer-deformed curve computed from Eq.~(53) (since $s_+/s_\alpha\neq 1$) with $D_{\pm}=\sin(2\mu_{\pm}P_{\pm})/\mu_{\pm}$, 
$K$ as in Fig.~\ref{fig:alpha_compare}, and $C_0=-1/s_\alpha$. 
The undeformed curve uses the $s_\alpha=0$ limit (Eq.~(57) with $s_{\pm}=0$), keeping the other parameters fixed. The polymer deformation 
leads to bounded oscillations, stabilizes the anisotropy dynamics at late times, preventing the monotonic divergence observed in 
the undeformed case.}
\label{fig:beta_plus_compare}
\end{figure}

Finally, we would like to emphasis that in the classical continuum limit \(\mu_i\to 0\) and \(s_i\to 0\), one recovers the usual Kasner-like linear behavior: \( \alpha \propto t,\ \beta_\pm \propto t\). Also, the presence of bounded trigonometric polymer functions can lead to bounded (oscillatory) or modified monotone behavior of the scale factor depending on parameters; however for the chosen gauge the dynamics reduces to elementary quadratures and closed-form expressions above.

In order to compare the standard Bianchi I dynamics with the polymer-deformed case, we have plotted the time evolution of the variables $\alpha(t)$, $\beta_+(t)$, and $\beta_-(t)$ in Figs.~\ref{fig:alpha_compare}--\ref{fig:beta_plus_compare} for typical values of the parameters. As these figures show clear qualitative differences between the standard Bianchi I cosmology 
and its polymer-deformed counterpart. The most striking feature is the suppression  of the unbounded growth in both the volume factor (through $\alpha$) and the anisotropy 
parameters ($\beta_+$, $\beta_-$). 

The volume variable $\alpha(t)$, exhibits a decreasing behavior with respect to the chosen time coordinate. This corresponds to a
contracting branch of the Bianchi~I dynamics, in which the spatial volume shrinks as time increases, eventually approaching a vanishing volume in the far future. Such a branch arises
naturally from Eq.~(47) when the combination of parameters $(s_\alpha,K)$, together with the chosen sign of $p_\alpha$ in the Hamiltonian constraint, yields a negative slope for $\alpha(t)$.

It is worth emphasizing that the same analytical framework also admits an expanding branch. By selecting the opposite sign for $p_\alpha$ or adjusting the parameters so that
$K$ changes sign, the model describes the usual cosmological scenario in which $\alpha(t)$ increases from $-\infty$ at an initial singularity (vanishing volume) to larger values as the
universe expands. This flexibility is a direct consequence of the time-reversal symmetry of the effective equations.

The anisotropy variables $\beta_{\pm}(t)$ follow the expected trends for the contracting case:
their amplitudes remain finite in the polymer--deformed model, while in the undeformed
classical dynamics they grow without bound near the collapse. For the anisotropy parameters $\beta_+$ and $\beta_-$, the polymer deformation prevents the monotonic divergence that occurs in the standard model. Instead, the anisotropies become bounded and undergo quasi-periodic oscillations. This suggests that the polymer modification acts as a stabilizing mechanism for the shape degrees 
of freedom of the Universe. Physically, such a bounded anisotropic behavior may imply that the Universe avoids certain types of shear-driven singularities present in the 
classical Bianchi~I dynamics. In the expanding branch, them roles of early and late times are interchanged, but the qualitative distinction between bounded
and unbounded anisotropy remains the same.

The combination of the classical polymerization with the deformed Poisson structure provides a richer phenomenology than either modification alone. In particular, the 
deformation parameter $s$, in the Poisson bracket introduces an additional scale that interacts non-trivially with the polymer length scale $\mu$. The interplay 
between these two parameters could lead to phase transitions in the cosmic dynamics, such as transitions between oscillatory and monotonic expansion regimes.

A complete analytical treatment of the singularity structure remains an open problem. Preliminary inspection of the numerical solutions indicates that the polymer-deformed 
model avoids certain curvature singularities by smoothing the behavior of the anisotropy variables, while still allowing for non-trivial oscillatory phases. 

\section{Asymptotic Behavior}

In the present model, the deformation of the Poisson structure is introduced through the relation (\ref{deform}), where the constants $s_i$ determine the strength and sign of the deformation in each canonical pair. So, their impact on the dynamics must therefore be analyzed directly from the 
solutions of the effective equations. Unlike the highly anisotropic and chaotic behavior of the Bianchi type~IX (Mixmaster) cosmology, 
the Bianchi type~I model exhibits a monotonic approach toward its asymptotic states, 
with the evolution of the volume variable $\alpha(t)$ and the anisotropy parameters $\beta_\pm(t)$ 
governed by Eqs.~(\ref{eq:alpha_solution_nonzero_s}) and (\ref{eq:beta_solution_general}) (or (\ref{eq:beta_solution_log}) in the marginal case $s_\pm=s_\alpha$). 
The exponential deformation $g_i(\alpha)$ enters these equations multiplicatively, 
modifying the effective rates of change of the variables. 
The sign of each $s_i$ plays a decisive role in determining whether the deformation 
enhances or suppresses anisotropies in a given asymptotic regime.

${\bullet}$ {\it{Large $\alpha$ limit (expanding branch)}}: In the limit $\alpha \to +\infty$, corresponding to the far future of an expanding branch, 
the deformation factors behave as

\begin{equation}
g_i(\alpha) \sim e^{s_i \alpha}.
\end{equation}
For $s_i>0$, the deformation grows exponentially, amplifying the corresponding dynamical term 
in the equations of motion. 
This can accelerate the decay of anisotropies if the amplified term acts as a restoring force, 
or enhance their growth if it acts in the opposite direction. 
For $s_i<0$, the deformation factor is exponentially suppressed at large volumes, 
effectively switching off its influence and allowing the dynamics to approach the undeformed 
(classical or purely polymerized) Bianchi~I behavior. 
The special case $s_i=0$ trivially reproduces the standard Poisson brackets.

${\bullet}$ {\it{Small $\alpha$ limit (contracting branch)}}: In the contracting branch, or in the past of an expanding branch, we have 
$\alpha \to -\infty$, corresponds to vanishing spatial volume. In this limit

\begin{equation}
g_i(\alpha) \sim e^{s_i \alpha} \to 
\begin{cases}
0 & \text{if } s_i>0,\\
+\infty & \text{if } s_i<0,
\end{cases}
\end{equation}
so that positive $s_i$ strongly suppress the corresponding dynamical contributions 
near the singularity, while negative $s_i$ blow them up. 
This implies that an appropriate choice of positive $s_\pm$ can bound the anisotropy variables 
$\beta_\pm(t)$ in the polymer--deformed model, in contrast with the unbounded growth found in 
the undeformed classical dynamics. 
Conversely, negative $s_\pm$ may enhance the anisotropic shear as the collapse is approached.

It should be noted that the polymer parameters $\mu_\alpha$ and $\mu_\pm$ enter the effective dynamics only through 
the combinations $\sin(2\mu_\alpha p_\alpha)/\mu_\alpha$ and 
$\sin(2\mu_\pm P_\pm)/\mu_\pm$, which appear in the constants 
$K$ and $D_\pm$ in the analytic solutions. 
As a result, they primarily rescale the overall rates of change of $\alpha(t)$ and $\beta_\pm(t)$ 
without altering the qualitative nature of the asymptotic behavior, which is instead governed 
by the signs and magnitudes of the deformation parameters $s_i$. 
Only for special values of $\mu$ leading to near-vanishing sine factors can the polymer scale 
qualitatively slow down the evolution even in the asymptotic regimes.

In summary, the exponential deformation studied here thus leads to the following qualitative behavior:
\begin{itemize}
\item For $s_i>0$, deformation effects are negligible at large volumes but can suppress anisotropies 
near the singularity.
\item For $s_i<0$, deformation effects are negligible near the singularity but can significantly 
modify the large-volume evolution.
\item The case $s_i=0$ reduces exactly to the undeformed polymerized model.
\end{itemize}
These trends are consistent with the solutions presented in previous section, where the choice $(s_\alpha>0,s_\pm=0)$ yields a contracting branch with bounded anisotropies.

\section{Conclusion}

In this work, we have investigated the dynamics of the Bianchi~I cosmological model under the combined 
effects of polymer quantization and an exponential deformation of the Poisson structure, given by 
$\{q_i,p_j\} = \delta_{ij} e^{s_i \alpha}$. 
Starting from a general Hamiltonian formulation, we implemented the polymerization 
in the canonical momenta and introduced the deformation through constant parameters $s_i$, which control 
the strength and sign of the exponential factor. 
Analytic solutions for the volume variable $\alpha(t)$ and the anisotropy parameters $\beta_\pm(t)$ were derived 
and compared to the undeformed polymerized dynamics.

The physical relevance of our results can be summarized as follows. 
Because the deformation of the Poisson algebra is explicitly volume-dependent, the effective dynamics 
exhibits qualitatively different features in the small- and large-volume regimes. 
For positive deformation parameters $s_i$, the anisotropy variables $\beta_\pm(t)$ remain bounded 
during the contracting phase, providing a mechanism to suppress shear that is absent in the undeformed 
classical model. At the same time, the slope of the logarithmic volume $\alpha(t)$ is reduced, indicating a slower 
approach to the singularity. 
Together, these effects suggest that exponential, volume-dependent deformations of the canonical 
structure may play a role in explaining how an initially anisotropic universe could evolve towards 
the near-isotropy observed today.

Our analysis, focused on the contracting branch of the model, shows that the deformation does not remove the 
initial singularity; the universe still evolves from a vanishing spatial volume at a finite past time. 
However, the deformation can significantly alter the approach to the singularity: for suitable $s_i$ values 
(especially $s_\pm>0$), the anisotropy variables $\beta_\pm(t)$ remain bounded, indicating a suppression of 
anisotropic shear that is absent in the undeformed case. 
In addition, the evolution of the volume variable $\alpha(t)$ is noticeably slower compared to the standard 
Bianchi~I model, reflecting the dynamical impact of the deformation on the effective expansion/contraction rates.

The asymptotic behavior, analyzed analytically and confirmed numerically, is governed primarily by the signs of 
the $s_i$ parameters. 
Positive $s_i$ suppress the corresponding dynamical contributions near the singularity while leaving the large-volume 
dynamics close to the undeformed case, whereas negative $s_i$ can enhance anisotropic effects at late times but 
remain negligible near the singularity. 
The polymer parameters $\mu_\alpha$ and $\mu_\pm$ act mainly as scale factors for the evolution rates and do not 
qualitatively change the asymptotic trends unless tuned to special values.

These findings illustrate that exponential deformations of the Poisson algebra offer a mechanism to control 
anisotropies in Bianchi-type cosmologies without altering the basic singularity structure. 
The framework presented here can be extended to other Bianchi models with nontrivial potentials, where the 
interplay between polymer effects and deformation parameters may lead to richer phenomenology, potentially 
including chaotic behavior in anisotropic dynamics.

It should be emphasized that the present analysis is primarily conceptual and is not directly 
compared with current cosmological data such as BAO, Pantheon, or Hubble measurements. 
A confrontation with observations would require extending the present formalism to the nearly 
isotropic limit, possibly with matter couplings or perturbative corrections, so that effective 
expansion histories can be extracted and contrasted with data. 
We regard this as an important direction for future work, beyond the scope of the current paper.

Finally, we note that a direct comparison with present-day cosmological observations such as 
BAO, Pantheon, or Hubble measurements is not feasible within the purely anisotropic, 
vacuum Bianchi~I framework studied here. 
Such a comparison would require extending the model to include matter sources and to 
examine the nearly isotropic limit, so that an effective expansion history can be confronted 
with data. 
While this lies beyond the scope of the present paper, we regard it as a natural and important 
direction for future research.

\end{document}